\documentclass{article}
\usepackage{arxiv}
\usepackage[utf8]{inputenc}
\usepackage[T1]{fontenc}
\usepackage{courier}
\usepackage{amsmath,amssymb,amsthm}
\usepackage{cite}
\usepackage{array,multirow,booktabs,graphicx}
\usepackage{algorithm,algpseudocode,flafter}
\usepackage{microtype}
\usepackage{xurl}
\usepackage{needspace}
\usepackage{etoolbox}
\usepackage{eso-pic}
\usepackage[hidelinks,unicode]{hyperref}

\renewcommand{\shorttitle}{CALM-5G: Event-Level Conformal Risk Control}
\renewcommand{\headeright}{}
\renewcommand{\undertitle}{}
\makeatletter
\patchcmd{\@maketitle}{\textsc{\undertitle}\\}{}{}{\PackageError{preprint}{Could not remove the preprint title line}{}}
\makeatother
\newcommand{\ieeesubmissionnotice}{%
  \begingroup
  \setlength{\fboxsep}{7pt}%
  \setlength{\fboxrule}{0.4pt}%
  \fbox{\begin{minipage}{\dimexpr\textwidth-2\fboxsep-2\fboxrule\relax}
    \footnotesize\raggedright
    This work has been submitted to the IEEE for possible publication. Copyright may be transferred without notice, after which this version may no longer be accessible.
  \end{minipage}}%
  \endgroup
}
\makeatletter
\patchcmd{\@maketitle}{\vskip 0.4in \@minus 0.1in \center{\@date} \vskip 0.2in}{\vskip 0.25in}{}{
  \PackageError{preprint}{Could not remove the date area}{}}
\newcommand{\bstctlcite}[1]{\@bsphack\if@filesw
  \immediate\write\@auxout{\string\citation{#1}}\fi\@esphack}
\newcommand{\keepheadingroom}[1]{\if@nobreak\else\Needspace{#1\baselineskip}\fi}
\makeatother
\pretocmd{\section}{\keepheadingroom{9}}{}{\PackageError{preprint}{Could not protect section headings}{}}
\pretocmd{\subsection}{\keepheadingroom{7}}{}{\PackageError{preprint}{Could not protect subsection headings}{}}
\pretocmd{\subsubsection}{\keepheadingroom{6}}{}{\PackageError{preprint}{Could not protect subsubsection headings}{}}

\title{CALM-5G: Event-Level Conformal Risk Control for Predictive Service-Assurance Alarms}
\author{\href{https://orcid.org/0009-0009-6006-7120}{\includegraphics[height=1.8ex]{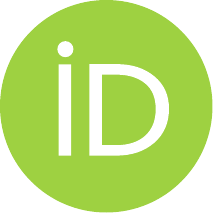}\hspace{0.4em}\"{O}mer~Faruk~\.{I}\c{s}canl\i{}}\\
\normalfont Siemens A.\c{S}., Istanbul, T\"{u}rkiye\\
\normalfont\texttt{faruk.iscanli@siemens.com}\\
\normalfont\small ORCID: \href{https://orcid.org/0009-0009-6006-7120}{0009-0009-6006-7120}
\And
\href{https://orcid.org/0000-0002-6356-452X}{\includegraphics[height=1.8ex]{orcid.pdf}\hspace{0.4em}Ya\u{g}mur~Sabucu~Sandal}\\
\normalfont Siemens A.\c{S}., Istanbul, T\"{u}rkiye\\
\normalfont\texttt{yagmur.sabucu@siemens.com}\\
\normalfont\small ORCID: \href{https://orcid.org/0000-0002-6356-452X}{0000-0002-6356-452X}}
\date{}

\hypersetup{
pdftitle={CALM-5G: Event-Level Conformal Risk Control for Predictive Service-Assurance Alarms},
pdfauthor={Ömer Faruk İşcanlı and Yağmur Sabucu Sandal},
pdfsubject={Manuscript submitted for review to IEEE Transactions on Network and Service Management},
pdfkeywords={5G, alarm management, concept drift, conformal prediction, conformal risk control, network management, service assurance, service-level objectives}
}

\begin{document}
\bstctlcite{CALMBSTcontrol}
% A first-page overlay occupies no space in the title or body layout.
\AddToShipoutPictureFG*{%
  \AtPageUpperLeft{%
    \put(\LenToUnit{1in},-\LenToUnit{24pt}){%
      \makebox(0,0)[lt]{\ieeesubmissionnotice}%
    }%
  }%
}
\maketitle

\begin{abstract}
Predictive service assurance converts forecasts of key performance indicators
into alarms, and that conversion decides whether the assurance function is
deployable. We present CALM-5G, an assurance layer that operates above any
forecasting backbone and issues alarms with distribution-free, finite-sample
guarantees. It calibrates on service-level violation events, scores violation
onset directly, selects its calibration scheme out of sample, and reports the
alarm duty cycle with every certified miss rate. Four corpora are evaluated
separately, namely a production cellular network, commercial 5G traces, an open
radio access network testbed and our 5G standalone testbed. Risk control on the
regression-derived score yields duty cycles of 0.24 to 0.91. Onset targeting
with cross-validated calibration reduces time-in-alarm by 56.3\% to 89.7\% on
three corpora, with no significant change on the fourth. The realised miss rate
stays within a two-standard-error tolerance on three corpora and within one
standard error of it on the fourth. Under label shift, pooled batch calibration
exceeds that tolerance structurally, by 10.9 standard errors on the production
corpus. Online calibration restores compliance. Calibration schemes cannot be
ranked on the split that fits their thresholds. Miss rate and duty cycle are
certified jointly where within-entity dependence leaves sufficient calibration
units.
\end{abstract}


\section{Introduction}
\label{sec:intro}

Mobile network management is shifting from repairing faults
after they are reported towards acting before a degradation reaches the
subscriber. The control loops being deployed read performance measurements on a
fixed cadence and issue actions within it~\cite{marinova2024oran}, and service
assurance couples that monitoring with analytics~\cite{xie2019assurance}.
Forecasts of key performance indicators (KPIs) have become a standard input to
these loops~\cite{schiavo2026tes}. A forecast, however, is not yet a decision.
Before it can drive an action it must be reduced to an alarm, and that
reduction determines whether the assurance function is usable. An alarm raised
too late prevents nothing, and an alarm raised too often is disabled by the
operations team. Neither failure is visible in the accuracy of the predictor.

\begin{figure}[!t]
\centering
\includegraphics[width=\textwidth,height=165pt,keepaspectratio]{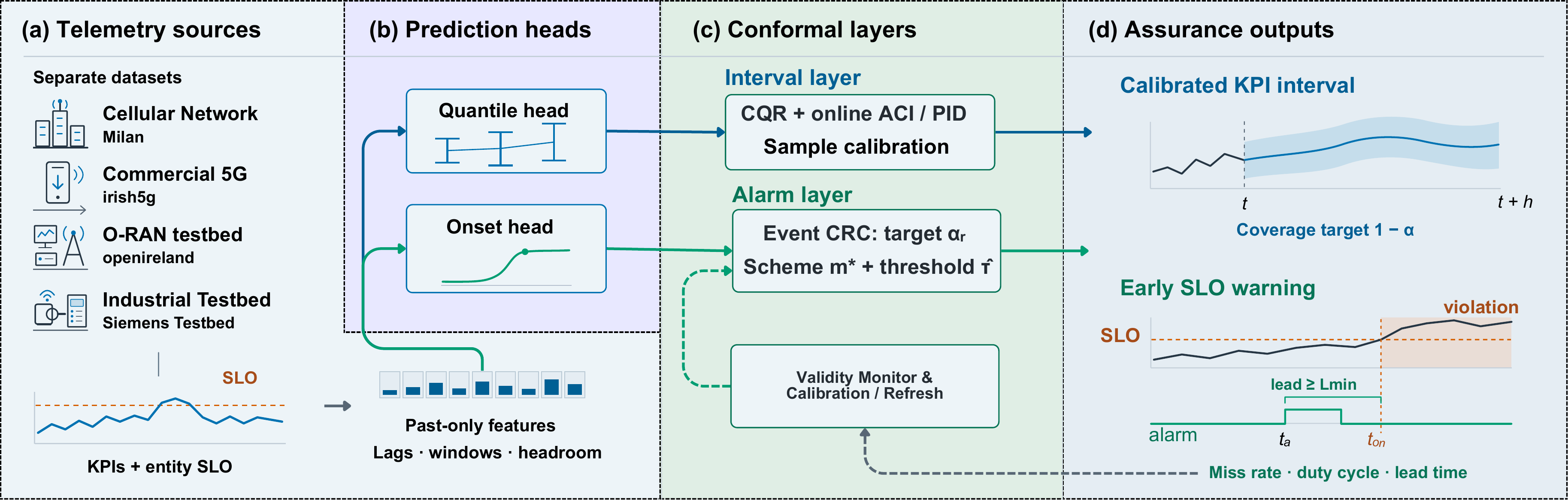}
\caption{CALM architecture. Four telemetry sources are evaluated separately. A
quantile forecaster and a classifier of violation onset are fitted on features
that use only past bins. The interval layer wraps the quantile head with
conformalised quantile regression and online adaptation. The alarm layer
thresholds a score with event-level conformal risk control at the nominal miss
rate $\alpha_{\mathrm{r}}$ under a calibration scheme $m^{\star}$ selected on
held-out folds, and a validity monitor requests recalibration or retraining. The
outputs are a calibrated KPI interval and an early warning that precedes the
event onset by at least $L_{\min}$.}
\label{fig:architecture}
\end{figure}

This paper treats that reduction as the object to be calibrated. It presents
CALM, an assurance layer that operates above an arbitrary forecasting backbone
and converts its output into alarms with distribution-free guarantees
(Fig.~\ref{fig:architecture}). CALM-5G denotes the instantiation for 5G
management evaluated here.

\subsection{Related Work}
\label{subsec:related}

The predictive component is mature. Learned models predict throughput from
radio and context measurements~\cite{narayanan2020lumos}, estimate latency
reliability from operator counters~\cite{conserva2026latency} and forecast KPIs
for anticipatory resource allocation~\cite{schiavo2026tes}. A parallel line
closes the loop at the radio access network (RAN) intelligent controller and
enforces service-level agreements~\cite{moro2023openran}. Both lines judge the
learned component by a population-averaged error or reward. Such summaries
describe behaviour over a test set and carry no commitment once the deployment
distribution moves. Where an end-to-end bound is established, it is supplied by
an analytical traffic model and not by the learned
predictor~\cite{xu2020slicing}. That non-stationarity is itself well
documented. Concept drift is a mature subject~\cite{lu2019drift}, model decay
after deployment has been characterised for networks~\cite{manias2023drift},
and recent drift detectors are evaluated on public corpora that are also used
here~\cite{ameur2025drift6g,fentazi2025aieurolens}. This literature establishes
when to retrain. It does not quantify how reliable the decisions of a model
remain while the shift is in progress.

Predictive uncertainty offers a route to per-decision trust, but not by itself
a guarantee. Ensemble estimates lose calibration under dataset
shift~\cite{lakshminarayanan2017ensembles,ovadia2019trust}. A comparison of
four families of interval estimators reported large, dataset-dependent
deviations from nominal coverage and identified conformal prediction as the
general repair~\cite{dewolf2023valid}. Conformal methods have accordingly been
adopted in wireless systems for demodulation, modulation classification and
channel prediction~\cite{cohen2023calibrating} and for the scheduling of
ultra-reliable traffic~\cite{cohen2023urllc}. Closer to network management,
they provide counterfactual KPI estimates at the RAN intelligent
controller~\cite{hou2025counterfactual} and context-dependent calibration by
meta-learning~\cite{yoo2026meta}. Online conformal risk control further
provides long-run guarantees on the false-negative rate of remote and edge
inference~\cite{zhu2025distributed,binucci2025lyapunov}. In all of these works
the calibrated quantity is defined per sample or per decision. None takes the
service-level violation event with a lead requirement as its unit, reports the
alarm duty that a miss-rate guarantee implies, or addresses how the calibration
scheme should be selected for a given deployment.

Three requirements therefore remain unmet where a forecast becomes an alarm.
First, the guarantee should be stated on the object the operator is accountable
for, namely the violation event together with the lead time an action requires.
Bins within an entity are strongly autocorrelated and inflate any effective
sample size computed from them. Second, controlling the miss rate is not a
specification. Risk control leaves the alarm rate free, and the admissible
threshold set always contains the always-alarm policy. The alarm budget must
therefore be reported alongside the miss rate and, where the data permit,
certified with it. Third, the calibration scheme cannot be assumed. Entities
differ in level by orders of magnitude, so the appropriate scheme is a property
of the deployment. It cannot, however, be chosen on the split used to fit its
threshold, because every candidate complies there by construction.

\subsection{Contributions}
\label{subsec:contrib}

CALM addresses these requirements. It separates an \emph{interval layer}, which
reports predictive uncertainty on the continuous service metric, from an
\emph{alarm layer}, which makes the assurance decision. Our contributions are
as follows.
\begin{itemize}
\item[\textbf{C1)}] A drift characterisation that precedes any claim of
adaptivity. Two months of a production cellular network contain two change
points, with a label shift that reduces the violation rate by almost an order
of magnitude from training to test. The baseline violation rate of our own 5G
standalone testbed varies by a factor of nine across identically scripted runs.
\item[\textbf{C2)}] CALM itself, which combines conformalised quantile
regression, online conformal updating, event-level conformal risk control and
abstention. We show that pooled batch calibration exceeds its nominal miss rate
\emph{structurally} under label shift, and that online calibration restores
compliance.
\item[\textbf{C3)}] A mapping of the framework onto the AI/ML lifecycle
management procedures of 5G management systems, together with an offline
actionability analysis that quantifies the lead time and alarm budget an
actuator would face.
\item[\textbf{C4)}] A four-corpus evaluation under a no-pooling contract. It
includes a testbed corpus with simultaneous application- and RAN-layer
observability, a layer ablation with a time-resolution control, and a compute
footprint measured with the onset classifier in the inference path.
\item[\textbf{C5)}] An alarm-rate result. Risk control applied to a
regression-derived score yields duty cycles of $0.24$ to $0.91$, and on two
corpora its miss rate is indistinguishable from that of a random alarm with the
same duty. Onset targeting with cross-validated calibration reduces
time-in-alarm by $56.3\%$ to $89.7\%$ on three of the four corpora, with
intervals excluding zero, and the reduction is not significant on the fourth.
The realised miss rate stays within a two-standard-error tolerance on three
corpora and within one standard error of it on the fourth.
\item[\textbf{C6)}] A selection and certification result of wider
applicability. Conformal calibration schemes cannot be ranked on the split used
to fit their thresholds, because the in-sample criterion admits every candidate
by construction. Both in-sample procedures examined here selected a scheme that
failed out of sample, and held-out scoring with deployment-matched folds
resolves the failure. Miss rate and alarm budget are further certified jointly
in $12$ of $20$ corpus--seed combinations under block dependence.
\end{itemize}

Section~\ref{sec:method} specifies the framework, Section~\ref{sec:setup} the
evaluation setup and Section~\ref{sec:results} the results, and
Section~\ref{sec:conclusion} concludes.

\section{Framework}
\label{sec:method}

This section specifies the system model, the two conformal layers, the
calibration procedure that makes them valid under the entity structure of
operational network data, and the end-to-end procedure that
Algorithm~\ref{alg:calm} summarises.

\subsection{Entities, Service-Level Events and Splits}
\label{subsec:system_model}

A managed network is partitioned into entities $i\in\mathcal{E}$, the unit at
which the assurance decision is taken, such as a cell, a flow or a measurement
session. Time is divided into bins of duration $\Delta$. For entity $i$ and bin
$t$ we observe a service metric $y_{i,t}$ and a feature vector $x_{i,t}$ built
only from bins up to $t-1$. The service-level objective (SLO) is a threshold
rule, and bin $(i,t)$ is \emph{violating} when
\begin{equation}
v_{i,t}=\mathbf{1}\!\left[\rho_i(y_{i,t})>\theta_i\right],
\label{eq:slo}
\end{equation}
where $\rho_i$ orients the metric so that an upper excursion is the violation
and $\theta_i$ is the threshold of entity $i$. For a throughput floor,
$\rho_i(y)=-y$. A \emph{violation event} is a maximal run of violating bins of
one entity, with onset $t_k^{\mathrm{on}}$ and duration $d_k$ bins. Events, not
bins, are the unit of assurance. An operator is concerned with the degradations
that arrived unwarned, not with the bins inside a degradation that was already
flagged.

Data are split into training, calibration and test parts according to the
intended deployment. Where entities persist, the split is chronological within
each entity. Where the model is to serve unseen entities, the split is grouped
so that no entity appears in two parts. This choice propagates to every
calibration decision below, so entity identity is treated as a first-class
object and not as one more categorical feature.

\subsection{Features and Prediction Heads}
\label{subsec:features}

Features comprise lags of the service metric and of the auxiliary counters,
rolling means and standard deviations over windows that end at $t-1$, and
time-of-day and day-of-week encodings. The onset head adds rolling quantiles
and their slopes, the interquartile range, and the headroom
$\theta_i-\rho_i(y_{i,t-1})$ with its slope. Lags and windows are specified in
bins so that the same specification transfers across the corpora of
Table~\ref{tab:corpora}. Lags reach back $144$ bins (one day) on Milan, $60$ on
Irish5G and $30$ on OpenIreland and the Testbed.

Two heads are fitted on the training part with gradient-boosted trees ($300$
trees, learning rate $0.05$, $31$ leaves, deterministic single-thread fitting,
no tuning). The framework is agnostic to this choice, and the guarantees below
hold for any measurable score. For a miscoverage level $\alpha$, the
\emph{quantile head} estimates the $\alpha/2$ and $1-\alpha/2$ quantiles, $\hat
q_{\mathrm{lo}}(x_{i,t})$ and $\hat q_{\mathrm{hi}}(x_{i,t})$, of $y_{i,t+h}$
under the pinball loss for a horizon of $h$ bins. The \emph{onset head} is a
classifier of
\begin{equation}
z_{i,t}^{(h)}=\mathbf{1}\!\left[\exists\,k:\ t<t_k^{\mathrm{on}}\le t+h\right],
\label{eq:onset}
\end{equation}
the indicator that a \emph{new} violation begins within the horizon. It is
trained on bins that are not themselves violating, because a system already in
violation cannot be warned of that violation, and whose horizon lies inside the
record. The positive class is reweighted by the class ratio, and the output
$r_{i,t}\in[0,1]$ is used as a score. Where the SLO rule smooths the metric
over $W$ bins we require $h\ge W$, so that no part of the regression target
$y_{i,t+h}$ is observable when the prediction is made.

\subsection{Interval Layer}
\label{subsec:interval}

The interval layer wraps the quantile head with conformalised quantile
regression (CQR)~\cite{romano2019cqr}. With calibration scores $E_j=\max\{\hat
q_{\mathrm{lo}}(x_j)-y_j,\;y_j-\hat q_{\mathrm{hi}}(x_j)\}$ and $\hat E$ their
$\lceil(n+1)(1-\alpha)\rceil$-th smallest value, the interval
\begin{equation}
\mathcal{C}(x)=\bigl[\hat q_{\mathrm{lo}}(x)-\hat E,\ \hat q_{\mathrm{hi}}(x)+\hat E\bigr]
\label{eq:cqr}
\end{equation}
covers a new sample with probability at least $1-\alpha$ when calibration and
test samples are exchangeable. Operational telemetry is not stationary, so the
framework also streams two online variants per entity. Adaptive conformal
inference (ACI)~\cite{gibbs2021aci} updates the working level as
\begin{equation}
\alpha_{t+1}=\alpha_t+\gamma\,(\alpha-\mathrm{err}_t),\qquad
\mathrm{err}_t=\mathbf{1}\!\left[y_t\notin\mathcal{C}_t\right],
\label{eq:aci}
\end{equation}
where $\mathcal{C}_t$ is \eqref{eq:cqr} with $\hat E$ replaced by the
$(1-\alpha_t)$ quantile of the last $1000$ scores, $\gamma=0.02$, and
$\alpha_t$ is clipped to $[0.001,0.999]$. Without clipping, ACI carries a
distribution-free bound on the long-run miscoverage frequency, so for the
clipped variant we rely on its measured coverage. A conformal
proportional-integral-derivative (PID) controller~\cite{angelopoulos2023pid}
augments the proportional term of \eqref{eq:aci} with an integrator over the
realised coverage error, which is better matched to telemetry with trends and
seasonality. Table~\ref{tab:interval} compares all constructions.

\subsection{Alarm Layer}
\label{subsec:alarm}

An alarm is raised when a score $s_{i,t}$ reaches a threshold,
$a_{i,t}(\tau)=\mathbf{1}[s_{i,t}\ge\tau]$. In the reference configuration the
score is the \emph{onset score} $s_{i,t}=r_{i,t}$. The layer is score-agnostic.
As a comparator we retain the \emph{regression score}, which is the upper end
of the CQR interval after orientation by $\rho_i$ and corresponds to the
conventional forecast-then-threshold construction.

Event $k$ of entity $i$ is \emph{detected} when an alarm is raised in the $h$
bins before its onset and at least $L_{\min}$ bins ahead of it, with
$L_{\min}=1$ bin throughout. With the event score
\begin{equation}
S_k=\max_{t_k^{\mathrm{on}}-h\,\le\,t\,\le\,t_k^{\mathrm{on}}-L_{\min}} s_{i,t},
\label{eq:eventscore}
\end{equation}
the miss loss is $\ell_k(\tau)=\mathbf{1}[S_k<\tau]$, which is non-decreasing
in $\tau$. For a nominal event false-negative rate (FNR), or miss rate,
$\alpha_{\mathrm{r}}$, conformal risk control (CRC)~\cite{angelopoulos2024crc}
on the $n$ calibration events selects
\begin{equation}
\hat\tau=\max\Bigl\{\tau\in\{S_k\}_{k=1}^{n}:\
\frac{|\{k:S_k<\tau\}|+1}{n+1}\le\alpha_{\mathrm{r}}\Bigr\},
\label{eq:crc}
\end{equation}
the largest threshold that satisfies the corrected constraint, and therefore
the one with the lowest alarm rate. If the calibration events and a test event
are exchangeable, then
\begin{equation}
\mathbb{E}\bigl[\ell_{n+1}(\hat\tau)\bigr]\le\alpha_{\mathrm{r}},
\label{eq:crc_guarantee}
\end{equation}
where the expectation is taken jointly over the calibration events and the test
event.

Two properties of \eqref{eq:crc_guarantee} delimit its scope and motivate the
remainder of this section. First, the guarantee is on an expectation and not on
the miss rate realised in a single test window. Compliance on a test split with
$n_{\mathrm{te}}$ events is therefore judged against the evaluation tolerance
\begin{equation}
\alpha_{\mathrm{r}}^{+}=\alpha_{\mathrm{r}}+2\sqrt{\alpha_{\mathrm{r}}(1-\alpha_{\mathrm{r}})/n_{\mathrm{te}}},
\label{eq:bound}
\end{equation}
which adds two event-level binomial standard errors to the nominal level.
Deviations are given in standard errors.

Second, the guarantee constrains the miss rate alone, since nothing in
\eqref{eq:crc} limits how often the alarm is raised. We therefore report the
\emph{duty cycle}, or time-in-alarm,
\begin{equation}
D(\tau)=\frac{1}{|\mathcal{D}|}\sum_{(i,t)\in\mathcal{D}}a_{i,t}(\tau)
\label{eq:duty}
\end{equation}
over all bins of the evaluated split alongside every miss rate, and read it
against two references (Fig.~\ref{fig:operating}). The first is the occupancy
\begin{equation}
D^{\star}=\frac{1}{|\mathcal{D}|}\sum_{k}\,(h+d_k)
\label{eq:floor}
\end{equation}
of an oracle that raises the alarm $h$ bins before each onset and holds it
until the event ends, counted per event. $D^{\star}$ is set by the event rate
and the horizon, and duty cycles are reported as multiples of it. The second
reference is a duty-matched \emph{chance level}, defined as the miss rate of
the same alarm sequence after an independent circular shift inside every
entity, averaged over $200$ draws. The shift preserves the duty and the
autocorrelation of the alarm and removes its alignment with the events. It
therefore separates alarm timing from alarm occupancy.

An abstention option exists for decisions close to the boundary. When the
margin $|s_{i,t}-\hat\tau|$ falls below a budgeted quantile of the calibration
margins, the decision is deferred to the operator. Abstained bins and the
retained fraction are reported separately.

\subsection{Calibration Schemes and Their Selection}
\label{subsec:calibration}

A single $\hat\tau$ is valid marginally over entities but can be poor
conditionally when entity levels span orders of magnitude. It saturates on
high-level entities, which then alarm almost continuously, while the events of
low-level entities are systematically missed. Whether a conditional scheme is
warranted is an empirical question about a deployment. CALM therefore supports
five schemes among which the deployment selects, all of which reduce to a
threshold rule on fixed heads.
\begin{itemize}
\item[\emph{M0}] \emph{Pooled batch.} One $\hat\tau$ from \eqref{eq:crc} over
all calibration events.
\item[\emph{M1}] \emph{Online.} Starting from the pooled threshold, the
threshold is updated after each event that has ended as
\begin{equation}
\tau_{k+1}=\tau_k-\eta\,(e_k-\alpha_{\mathrm{r}}),\qquad
e_k=\mathbf{1}[S_k<\tau_k],
\label{eq:online}
\end{equation}
with $\eta$ equal to $0.3$ standard deviations of the calibration event scores.
Events of all entities are processed in the order in which they end.
\item[\emph{M2}] \emph{Covariate-conditional.} Bins are grouped by quantiles of
a covariate that is known at prediction time and defined for unseen entities,
namely the trailing level of the service metric. Equation~\eqref{eq:crc} is
then applied per group, as in group-conditional conformal
prediction~\cite{melki2023groupcond}. The number of groups is the largest of
$3$, $5$ and $10$ that leaves at least $30$ calibration events in each.
\item[\emph{M3}] \emph{Online conditional.} M1 inside the groups of M2.
\item[\emph{M4}] \emph{Entity-standardised.} Scores are centred and scaled per
entity before a pooled threshold is applied.
\end{itemize}
Summing \eqref{eq:online} over $T$ events gives the identity
\begin{equation}
\frac{1}{T}\sum_{k=1}^{T}e_k-\alpha_{\mathrm{r}}=\frac{\tau_1-\tau_{T+1}}{\eta\,T},
\label{eq:identity}
\end{equation}
which holds for any sequence of events. The online miss rate therefore
converges to $\alpha_{\mathrm{r}}$ at rate $1/T$ as long as the threshold
remains bounded, without any distributional assumption (Fig.~\ref{fig:milan}).
This is the event-level counterpart of the long-run guarantees established for
per-decision losses in~\cite{zhu2025distributed,binucci2025lyapunov}.

The scheme is itself a modelling decision and must be chosen without the test
part. It cannot be chosen by scoring the schemes on the calibration part,
because \eqref{eq:crc} fits the threshold there and every scheme then attains a
miss rate of approximately $\alpha_{\mathrm{r}}$ by construction. CALM
therefore selects by cross-validation \emph{inside} the calibration part, with
folds that mirror the deployment split. Folds are grouped by entity where the
target is unseen entities. Where entities persist, folds are contiguous
forward-in-time blocks, each scored by thresholds fitted on earlier blocks
only.

With $\bar D_m$ the mean held-out duty of scheme $m$ over repetitions, the rule
is
\begin{equation}
m^{\star}=\arg\min_{m\in\mathcal{A}}\bar D_m ,
\label{eq:selection}
\end{equation}
where the admissible set $\mathcal{A}$ contains the schemes that satisfy three
conditions. The held-out miss rate must lie within the tolerance
\eqref{eq:bound} on at least four of five repetitions, and the largest held-out
duty must not exceed twice the median. In addition, no held-out fold may show a
duty below $0.01$ together with a miss rate above $0.5$, which rejects schemes
that leave individual entities silent. A fold must hold at least $30$ events
and two entities, and corpora whose calibration part cannot supply them fall
back to a single-repetition selection.

\subsection{Joint Miss-Rate and Alarm-Budget Certification}
\label{subsec:joint}

Where an explicit alarm budget is required, the threshold rule is replaced by a
hysteretic state machine with configuration
$\lambda=(u,w,c,\varrho,\tau_{\mathrm{on}},\tau_{\mathrm{off}})$. The alarm
starts when $u$ of the last $w$ bins satisfy $s_{i,t}\ge\tau_{\mathrm{on}}$,
ends after $c$ consecutive bins below $\tau_{\mathrm{off}}$, and no new
notification is issued for $\varrho$ bins after it ends. A notification is
emitted only when an alarm starts, so the notification count and the
time-in-alarm are distinct quantities. A grid $\Lambda$ of $648$ configurations
is searched, and Learn-then-Test (LTT)~\cite{angelopoulos2025ltt} supplies the
multiplicity control. For a risk level $\alpha_{\mathrm{J}}$, an alarm budget
$\beta$ and each $\lambda$, it tests the nulls
\begin{equation}
H_1(\lambda):\ R(\lambda)>\alpha_{\mathrm{J}},\qquad
H_2(\lambda):\ D(\lambda)>\beta
\label{eq:ltt}
\end{equation}
on the miss rate $R$ and the duty $D$. The Hoeffding--Bentkus $p$-values are
combined as $\max\{p_1,p_2\}$ and rejected under a Bonferroni correction over
the grid at $\delta=0.1$. For independent calibration units and a fixed pair
$(\alpha_{\mathrm{J}},\beta)$, every surviving configuration then satisfies
both constraints with probability at least $1-\delta$. Among the survivors
whose median calibration lead is at least $L_{\mathrm{J}}$, CALM adopts the
configuration with the smallest calibration duty. Minimising the notification
count instead would favour an alarm that never ends.

The budget is not fixed in advance. Two budgets are admissible, the tighter
$2D^{\star}_{\mathrm{cal}}$ and the looser calibration duty of the regression
baseline, and each is paired in turn with the risk levels $0.05$, $0.10$ and
$0.20$. The procedure evaluates these pairs in order on calibration data and
stops at the first that admits a survivor, so a corpus certifies at the
tightest budget and risk level that its event process allows. The risk levels
are nested and the $p$-values are non-increasing in the level, so for a given
budget the guarantee holds simultaneously over the three levels. The level is
reported with every certificate.

Alarm indicators are strongly autocorrelated (Fig.~\ref{fig:dependence}). The
duty $p$-value is therefore computed on the means of contiguous within-entity
blocks of $B$ bins. The block length $B$ is the first lag at which the
calibration autocorrelation of the alarm indicator falls below $0.1$, and the
number of blocks is taken as the effective sample size. The miss-rate $p$-value
treats calibration events as units. Results with one unit per bin and one per
entity bracket this choice, and a set that is non-empty only under the per-bin
model carries no duty certificate.

\subsection{Lifecycle Monitoring and Procedure}
\label{subsec:lifecycle}

Validity is monitored during operation from quantities that the framework
already produces. A monitor combines the rolling coverage gap of the interval
layer with a change-detection statistic on the stream of conformity scores and
moves between \textsc{stable}, \textsc{degraded} and \textsc{recovering}
states. These states request, in order of cost, recalibration, retraining and
rollback. Recalibration is attempted first because it is less costly and
suffices when drift is confined to the score distribution. The states and
transitions correspond to the AI/ML lifecycle management procedures
standardised for 5G management systems. Model training, testing and deployment
map onto 3GPP TS~28.105, the retraining and rollback transitions onto the
closed control loop management of TS~28.567, and the measurement path onto the
analytics accuracy information of TS~23.288. In Algorithm~\ref{alg:calm}, every
step before the online loop reads the training and calibration parts only.

\begin{algorithm}[!t]
\footnotesize
\caption{CALM construction, scheme selection, optional budget certification
and online operation}
\label{alg:calm}
\begin{algorithmic}[1]
\Statex \textbf{Input:} telemetry $\{(x_{i,t},y_{i,t})\}$, thresholds
        $\{\theta_i\}$, levels $\alpha$, $\alpha_{\mathrm{r}}$, horizon $h$
\State Label violations by \eqref{eq:slo} and onset targets by
       \eqref{eq:onset}, then split by deployment
\State Fit the quantile head and the onset head on the training part
\State $\hat E\leftarrow$ CQR quantile on the calibration part, and start
       \eqref{eq:aci} per entity
\For{each scheme $m\in\{\mathrm{M0},\dots,\mathrm{M4}\}$ and each fold $F$}
  \State Fit the thresholds of $m$ by \eqref{eq:crc} on the fitting folds
  \State Record miss rate and duty \eqref{eq:duty} of $m$ on $F$
\EndFor
\State $m^{\star}\leftarrow$ \eqref{eq:selection}, refitted on the whole
       calibration part
\If{an alarm budget $\beta$ is required}
  \State $B\leftarrow$ first lag with calibration autocorrelation below $0.1$
  \State Test \eqref{eq:ltt} for all $\lambda\in\Lambda$ on events and blocks
  \State Adopt the survivor with lead $\ge L_{\mathrm{J}}$ and the smallest
         calibration duty
\EndIf
\For{each arriving bin $(i,t)$}
  \State Emit $\mathcal{C}_t(x_{i,t})$ and update $\alpha_t$ by \eqref{eq:aci}
         when the label arrives
  \State Raise the alarm if $s_{i,t}\ge\tau$, or by the state machine
         $\lambda^{\star}$
  \State When an event ends, update $\tau$ by \eqref{eq:online} if
         $m^{\star}$ is online
  \State Update the monitor and recalibrate, retrain or roll back on demand
\EndFor
\end{algorithmic}
\end{algorithm}

\subsection{Assumptions and Scope}
\label{subsec:scope}

The guarantees rest on stated conditions. The coverage of \eqref{eq:cqr} and
the risk bound \eqref{eq:crc_guarantee} hold under exchangeability of
calibration and test units, and \eqref{eq:crc_guarantee} is an expectation over
the calibration draw. The online update \eqref{eq:online} satisfies
\eqref{eq:identity} for every event sequence without a distributional
condition, provided that each event outcome is observed. The joint certificate
assumes independent blocks and events. These guarantees hold for each scheme as
a fixed procedure. Scheme selection is empirical, and its outcome is validated
on the untouched test part. Operational telemetry departs from exchangeability,
and it is precisely for that reason that the online variants and the
event-level calibration unit are constitutive of the design and not refinements
of it. The framework observes and advises. The alarm stream that it produces is
the input to whatever control action a deployment exposes, and the outcome of a
particular actuation policy is outside its scope.

\section{Evaluation Setup}
\label{sec:setup}

\subsection{Corpora and Service-Level Rules}
\label{subsec:corpora}

Table~\ref{tab:corpora} lists the four corpora, which are never pooled. Each
has its own loader, SLO rule, bin size, feature windows and split, and no model
or threshold is shared.

\emph{Milan} is the Telecom Italia call-detail-record
dataset~\cite{barlacchi2015milan,telecomitalia2015milan}, which reports the
internet activity per grid cell in $10$-minute bins from 1~November 2013 to
1~January 2014. We use $200$ cells, the $150$ busiest on the first day and $50$
drawn at random from the rest. A bin is violating when the activity exceeds the
$95$th percentile of the cell over the training period. The split is
chronological, so the test period, 20~December to 1~January, coincides with the
Christmas holiday. This corpus contributes scale and a longitudinal shift from
a production network, while the other three supply radio realism.

\emph{Irish5G} contains $64$ traces of a commercial 5G network sampled at
$1$\,s~\cite{raca20205g}. \emph{OpenIreland} comprises $20$ runs of an
\mbox{O-RAN} testbed at $100$\,ms with benign and attack
traffic~\cite{xavier2024oran}, where the attack rows are held out as a stress
partition. For both corpora the SLO is a sustained throughput floor. A bin is
violating when the $10$-bin mean downlink throughput falls below a floor of $3$
and $11.2$\,kbit/s, which yields violation rates of $4.8\%$ and $5.9\%$. Both
corpora are split by trace or run.

\begin{table}[!t]
\renewcommand{\arraystretch}{1.12}
\caption{Corpora, Service-Level Rules and Splits. Ranges Are Over the Five
Seeds}
\label{tab:corpora}
\centering
\footnotesize
\setlength{\tabcolsep}{3.2pt}
\begin{tabular}{@{}lcccc@{}}
\toprule
 & Milan & Irish5G & OpenIreland & Testbed \\
\midrule
Bin $\Delta$ & 10\,min & 1\,s & 100\,ms & 1\,s \\
Entities & 200 cells & 64 traces & 20 runs & 18 runs \\
Metric & activity & throughput & throughput & $p_{95}$ delay \\
SLO rule & $>$ cell $p_{95}$ & $<3$\,kbit/s & $<11.2$\,kbit/s & $>50$\,ms \\
Split & chronol. & by trace & by run & by run \\
Train/cal./test & 60/20/20\% & 38/13/13 & 12/4/4 & 8/6/4 \\
Horizon $h$ (bins) & 1 & 10 & 10 & 10 \\
Violation rate & 0.036 & 0.048 & 0.059 & 0.037 \\
Test bins ($10^{3}$) & 357 & 19--26 & 59--265 & 4.1--5.0 \\
Test events & 654 & 167--227 & 363--2024 & 96--146 \\
Event length (bins) & 3.1 & 4.4--8.0 & 7.5--11.0 & 1.4--1.6 \\
\bottomrule
\end{tabular}
\end{table}

\subsection{Testbed}
\label{subsec:labsetup}

The fourth corpus was recorded on our own 5G standalone testbed. It consists of
an OpenAirInterface gNB and one user equipment (UE) on USRP~B210
software-defined radios with a containerised core. The carrier spans $24$
physical resource blocks at $30$\,kHz subcarrier spacing with seven downlink
and two uplink slots per period. The uplink carries live telemetry of a servo
drive, which a gateway on the UE host reads from a Siemens S7-1200 programmable
logic controller and publishes over MQTT at about $13$\,kbit/s. The evaluation
therefore includes measurements from physical industrial equipment alongside
the public corpora.

On a capacity-limited uplink, degradation appears as queueing delay and not as
loss, so the SLO is placed on delay. A $1$-s bin is violating when the $95$th
percentile of the round-trip time of a $50$-Hz echo probe exceeds $50$\,ms. The
probe carries four timestamps, so no clock synchronisation between the hosts is
required. Impairment episodes are created by a rate-controlled UDP load that
rises linearly from a base of $2.5$\,Mbit/s to a peak of $3.8$--$6.0$\,Mbit/s,
is held and then released. A ramp is used because a step has no precursor and
would make any measured lead time an artefact of the schedule. RAN counters
arrive every $1.28$\,s and are forward-filled by at most two bins, with the
number of stale bins exposed as a feature.

Eighteen runs ($6.1$\,h) are retained under fixed loader rules on probe loss,
baseline stability, RAN coverage and log alignment. The violation rate of their
baseline phases, which are scripted identically across runs, ranges from
$0.0096$ to $0.0862$. This factor of nine is uncontrolled and not induced, so
the Testbed most directly stresses exchangeability between calibration and
test. It is also the only corpus in this evaluation with application-layer
delay and RAN counters observed on the same link, at the scale of one link, one
UE and one cell.

\subsection{Protocol}
\label{subsec:protocol}

All layers use $\alpha=0.10$ and $\alpha_{\mathrm{r}}=0.05$, and each corpus
has a primary horizon (Table~\ref{tab:corpora}). Results are reported over five
seeds, which change the learner and, on the grouped corpora, the entity split.
Intervals written as mean $[\mathrm{lo},\mathrm{hi}]$ are $95\%$ bootstrap
intervals over the seeds, and ranges give the minimum and maximum. Diagnostics
that isolate a single mechanism are run on one seed and are identified as such.
Within a seed, no test data enter feature construction, model fitting or
threshold fitting, and the Milan thresholds $\theta_i$ are computed on the
training period. Re-running one configuration per corpus reproduces all $350$
metric values exactly.

The interval layer is compared with the uncalibrated quantile head, split
conformal prediction around a point forecast, a Gaussian interval from the
training residuals and the spread of a ten-member bagged ensemble. The alarm
layer is compared against the regression score with pooled CRC, which is the
conventional regression-then-threshold construction, and against the onset
score with pooled CRC, which changes only the score (Table~\ref{tab:arms}). The
\emph{adopted} configuration is the onset score with the scheme that
\eqref{eq:selection} returns, and Table~\ref{tab:selection} lists all schemes.
Where the adopted calibration is pooled, on Irish5G and the Testbed, the two
onset configurations coincide and the whole gain is attributable to the score.
Events are identified inside each split, and an event that begins in the first
bin of a test block cannot be warned and is conservatively counted as a miss.
The lead requirements $L_{\mathrm{J}}$ of the joint certification are $1$, $5$,
$5$ and $3$ bins, and Table~\ref{tab:dependence} lists its levels.

\section{Results}
\label{sec:results}

\subsection{Interval Layer and Adaptation Under Drift}
\label{subsec:res_interval}

\begin{table}[!t]
\renewcommand{\arraystretch}{1.1}
\caption{Interval Layer at the Primary Horizon. Coverage of Nominal $90\%$
Intervals as the Mean Over Five Seeds, With the Mean Width Relative to Static
CQR in Parentheses}
\label{tab:interval}
\centering
\footnotesize
\setlength{\tabcolsep}{2.8pt}
\begin{tabular}{@{}lcccc@{}}
\toprule
 & Milan & Irish5G & OpenIreland & Testbed \\
\midrule
Quantile head & 0.920 (1.02) & 0.846 (0.95) & 0.888 (1.00) & 0.827 (0.93) \\
CQR, static & 0.878 (1.00) & 0.909 (1.00) & 0.896 (1.00) & 0.914 (1.00) \\
ACI, online & 0.900 (0.96) & 0.900 (1.13) & 0.900 (1.04) & 0.900 (1.05) \\
Conformal PID & 0.953 (1.12) & 0.934 (1.24) & 0.941 (1.13) & 0.931 (1.06) \\
Split conformal & 0.954 (1.46) & 0.884 (2.39) & 0.900 (1.24) & 0.900 (1.17) \\
Gaussian & 0.988 (2.45) & 0.894 (1.61) & 0.926 (1.40) & 0.940 (1.51) \\
Bagged ensemble & 0.108 (0.06) & 0.125 (0.16) & 0.045 (0.05) & 0.403 (0.38) \\
\bottomrule
\end{tabular}
\end{table}

At the primary horizon the online conformal layer attains its nominal level on
all four corpora (Table~\ref{tab:interval}), with ACI reaching a coverage of
$0.8995$, $0.9002$, $0.8995$ and $0.9005$ against a target of $0.90$. Static
CQR deviates slightly (\mbox{$0.878$--$0.914$}), the PID controller over-covers
($0.931$--$0.953$), and split conformal prediction over-covers on Milan with
intervals $46\%$ wider than those of CQR. The non-conformal references bracket
the failure modes that the conformal layer excludes. The Gaussian interval
over-covers at $0.988$ on Milan, whereas the bagged ensemble collapses to
$0.108$, which confirms that an ensemble spread is not a calibrated
interval~\cite{lakshminarayanan2017ensembles,ovadia2019trust}. On a random
split of the same data, where exchangeability holds by construction, CQR covers
$0.889$--$0.904$, so its deviations are attributable to shift. Validity does
not come at the expense of sharpness, since the quantile head attains $0.32$,
$0.24$, $0.24$ and $0.49$ times the pinball loss of the persistence baseline.

Drift is characterised before adaptivity is claimed. The two months of the
production network contain two change points, with shifts of $139.6$ and
$159.4$ activity units affecting $70.8\%$ and $75.0\%$ of the cells. The
violation rate falls from $0.050$ in training to $0.026$ in calibration and
$0.006$ in test, the label shift that the alarm layer must withstand. In the
change-point windows the coverage of the static layer deviates from nominal by
up to $0.0707$, against $0.0026$ in calibration, while ACI maintains $0.90$.
Day by day and over all seeds, static CQR covers between $0.826$ and $0.926$
and split conformal prediction between $0.886$ and $0.977$, while ACI stays
within $0.893$--$0.917$.

The OpenIreland stress partition adds a labelled abrupt shift, with all models
fitted on benign data only. ACI and PID recover there in $4.5$ and $0$ bins,
against $63.75$ bins for the static layer, at a width inflation of $1.21$ to
$1.27$. Drift-triggered lifecycle management dominates periodic retraining
where a continuous timeline exists. It attains a coverage of $0.9508$ with
$0.6$ retrains against $0.6459$ with $53$ on Milan, and $0.8537$ with $0.4$
against $0.7456$ with $8$ on the Testbed.

\subsection{Alarm Rate Under Risk Control}
\label{subsec:res_alarmrate}

\begin{figure}[!t]
\centering
\includegraphics[width=0.56\linewidth]{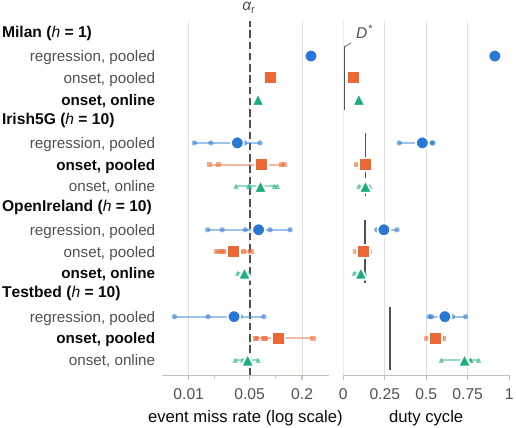}
\caption{Miss rate and duty cycle of three configurations at the primary
horizon. Small marks are seeds, large marks their mean, and bold labels the
adopted configuration. The dashed line is the nominal miss rate and the vertical
ticks the reference occupancy $D^{\star}$ of each corpus.}
\label{fig:operating}
\end{figure}

With the regression score, pooled CRC at $\alpha_{\mathrm{r}}=0.05$ yields duty
cycles of $0.914$, $0.477$, $0.245$ and $0.613$ (Table~\ref{tab:arms},
Fig.~\ref{fig:operating}). Since \eqref{eq:crc} returns the largest admissible
threshold, these are the lowest duty cycles that this score admits at that
level.

Two mechanisms are separable. On Irish5G, OpenIreland and the Testbed the mean
interval width is $135.4$, $5.65$ and $2.76$ times the distance from the benign
operating point to the SLO threshold. No threshold on that score can therefore
be both sensitive and selective. On Milan the ratio is $0.166$ and the cause is
scale. Cell thresholds span $4.83$ to $4111.94$ activity units, and a pooled
threshold in those units holds the alarm active for more than $99\%$ of the
time in \mbox{$167$--$169$} of the $200$ cells. A single-seed diagnostic that
removes the level of each cell from the score, with the threshold left pooled,
reduces the duty from $0.914$ to $0.171$ at an event FNR of $0.055$. The
binding defect on Milan is therefore scale and not the score.

\begin{table}[!t]
\renewcommand{\arraystretch}{1.12}
\caption{Alarm Layer at the Primary Horizon. Dev. Is the Mean Deviation of the
Event FNR From the Tolerance in Standard Errors, With Negative Values Inside,
and Chance Is the Miss Rate After Random Circular Shifts. Bold Marks the
Adopted Configuration}
\label{tab:arms}
\centering
\footnotesize
\setlength{\tabcolsep}{2.6pt}
\begin{tabular}{@{}llccccc@{}}
\toprule
Corpus & Score, calibration & Duty & Event FNR & Tol. & Dev.\,(SE) & Chance \\
\midrule
\multirow{3}{*}{Milan}
 & regr., pooled & 0.9138 & 0.2529 & \multirow{3}{*}{0.0670} & $+10.9$ & 0.30 \\
 & onset, pooled & 0.0644 & 0.0872 & & $+1.8$ & 0.72 \\
 & \textbf{onset, online} & \textbf{0.0938} & \textbf{0.0627} & & $\mathbf{-0.5}$ & \textbf{0.64} \\
\midrule
\multirow{2}{*}{Irish5G}
 & regr., pooled & 0.4767 & 0.0362 & \multirow{2}{*}{0.0815} & $-3.5$ & 0.06 \\
 & \textbf{onset, pooled} & \textbf{0.1352} & \textbf{0.0680} & & $\mathbf{-0.8}$ & \textbf{0.17} \\
\midrule
\multirow{3}{*}{OpenIr.}
 & regr., pooled & 0.2449 & 0.0634 & \multirow{3}{*}{0.0671} & $-0.4$ & 0.55 \\
 & onset, pooled & 0.1221 & 0.0326 & & $-5.1$ & 0.78 \\
 & \textbf{onset, online} & \textbf{0.1058} & \textbf{0.0437} & & $\mathbf{-3.0}$ & \textbf{0.80} \\
\midrule
\multirow{2}{*}{Testbed}
 & regr., pooled & 0.6127 & 0.0331 & \multirow{2}{*}{0.0914} & $-3.6$ & 0.07 \\
 & \textbf{onset, pooled} & \textbf{0.5545} & \textbf{0.1068} & & $\mathbf{+0.6}$ & \textbf{0.19} \\
\bottomrule
\end{tabular}
\end{table}

The chance level quantifies the information in these alarms. On Irish5G and the
Testbed the regression-score alarm misses $3.6\%$ and $3.3\%$ of the events,
whereas its randomly shifted copy misses $5.8\%$ and $6.6\%$. The observed
value lies inside the central $95\%$ of the shifted draws on four and five of
five seeds, so the nominal level is met largely through occupancy.

On Milan, Irish5G and OpenIreland the adopted configuration stays within its
tolerance and reduces time-in-alarm by $89.7\%\,[89.3,90.3]$ on Milan,
$71.1\%\,[65.5,77.9]$ on Irish5G and $56.3\%\,[47.5,64.1]$ on OpenIreland, with
intervals excluding zero. Compliance is judged on the across-seed mean
deviation in standard errors and not on the point estimate
(Table~\ref{tab:arms}). On Milan and OpenIreland the adopted configuration
improves both the miss rate and the duty. On Milan its duty of $0.094$ also
lies below the $0.171$ of the level-corrected regression score. Its miss rate
lies below the chance level on every corpus, by a factor of $1.8$ on the
Testbed and of $2.5$ to $18$ elsewhere.

Duty cycles are read against the reference occupancy. The adopted alarm
operates at $1.0$ and $0.8$ times $D^{\star}$ on Irish5G and OpenIreland.
Multiples below one are attainable because an alarm may start later than $h$
bins before the onset. On Milan the multiple is larger, at $12.6$, because the
low event rate of that corpus yields a low reference. On the Testbed, events
last $1.5$\,s on average and recur about every $40$\,s, so the reference
occupancy is the highest of the four, at $D^{\star}=0.28$, which limits the
achievable reduction. The reduction there is $7.3\%\,[-8.2,22.0]$, an interval
including zero, and the realised FNR lies within one standard error of the
tolerance.

The gain on the first three corpora persists at the other stored horizons. The
reduction is $92\%$ and $93\%$ on Milan at $h=3$ and $6$, $47\%$ and $57\%$ on
OpenIreland at $h=30$ and $100$, and $64\%$ and $31\%$ on Irish5G at $h=30$ and
$120$.

The ablations attribute the gain. Substituting the regression score raises the
duty, in absolute terms, by $0.820$, $0.342$ and $0.139$ on the first three
corpora, with intervals excluding zero. Substituting pooled calibration where
it is not the adopted scheme lowers the duty on Milan by $0.029$ but moves the
event FNR to $0.0872$, outside the tolerance, and raises the duty on
OpenIreland by $0.016$. Removing risk control, with a fixed threshold at the
calibration quantile that matches the onset base rate, gives duties of $0.002$,
$0.075$, $0.066$ and $0.190$ at event FNRs of $0.956$, $0.201$, $0.178$ and
$0.404$. A fixed threshold therefore yields a different operating point and not
a less costly alarm. Abstention acts on the retained decisions. On the
regression baseline, abstaining the most ambiguous $29.4\%$ of Milan decisions
lowers the event FNR from $0.2456$ to $0.0017$, and $9.7\%$ of Testbed
decisions from $0.0366$ to $0.0080$.

Under the adopted configuration the median lead inside the detection window is
$6$--$10$ bins on the grouped corpora and one bin on Milan. This corresponds to
$9$--$10$\,s on Irish5G, $10$\,s on the Testbed and $0.6$--$0.9$\,s on
OpenIreland. New alarms start $0.063$, $55$, $642$ and $168$ times per hour, a
rate that the state machine of Section~\ref{subsec:joint} reduces.

\subsection{Miss-Rate Compliance Under Label Shift}
\label{subsec:res_shift}

Because \eqref{eq:crc_guarantee} holds in expectation and not per test window,
compliance is judged on the across-seed mean and its deviation in standard
errors. On Milan, where the seeds share one chronological split, a bootstrap
over cells adds a sampling reference. The regression baseline realises an event
FNR of $0.2529\,[0.2523,0.2535]$ against a tolerance of $0.0670$, a deviation
of $+10.9$\,SE that is present on all five seeds. The violation is structural
and not sampling noise. The five cells in which the pooled threshold almost
never alarms hold $3\%$ of the calibration events but $19\%$ of the test
events, and they account for $121$ of the $165$--$166$ misses. During the
holiday, violations move to cells the pooled threshold does not reach.

Changing only the score gives $0.0872\,[0.0844,0.0905]$, still $+1.8$\,SE and
above the tolerance on all five seeds. Only online calibration returns the
realised FNR inside the tolerance, at $0.0627\,[0.0615,0.0639]$ and $-0.5$\,SE,
on every seed. The two contributions share a mechanism. The pooled threshold
that exceeds the tolerance under label shift also produces the excessive alarm
rate above, so one change addresses both.

\begin{figure}[!t]
\centering
\includegraphics[width=0.58\linewidth]{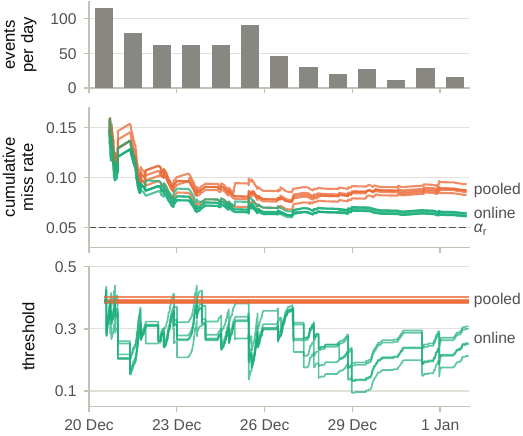}
\caption{Milan test period under the onset score, with one line per seed. The
top panel shows the events per day, the middle panel the cumulative miss rate
of the pooled and the online threshold against the nominal level (dashed), and
the bottom panel the two thresholds.}
\label{fig:milan}
\end{figure}

Fig.~\ref{fig:milan} follows the test period. Events per day fall from $116$ on
the first day to \mbox{$12$--$30$} in the last six days. The online threshold
adapts within the first day and remains below the pooled value for most of the
period. From the second day on, the online threshold misses
\mbox{$4.8$--$5.2\%$} of the events, in line with the nominal level, against
$7.2$--$8.6\%$ for the pooled threshold. Two effects account for the online
excess. Six events begin in the first test bin of their cell and cannot be
warned, which contributes $0.009$. Over the other $648$ events the online miss
rate is \mbox{$0.052$--$0.056$}, against \mbox{$0.074$--$0.085$} for the pooled
threshold, and identity \eqref{eq:identity} attributes the remainder to the net
drop of the online threshold. The reported miss rate uses the threshold in
force at each bin and differs from the loss in \eqref{eq:identity} by at most
two events. A bootstrap over cells gives $95\%$ intervals whose lower end is
$0.057$--$0.072$ for the pooled miss rate and $0.041$--$0.045$ for the online
one. The pooled threshold therefore exceeds the nominal level beyond what
cell-level sampling explains, whereas the online one does not.

The Testbed corroborates this in a control with the regression score at $h=5$.
Over $200$ random partitions of twelve runs into calibration and evaluation
halves, exchangeable at the run level by construction, the mean miss rate of
pooled CRC is $0.053$, in agreement with \eqref{eq:crc_guarantee}. Individual
windows scatter around that mean, and $21\%$ of the $400$ evaluations exceed
the batch tolerance, whereas none does under online calibration.

\subsection{Selecting the Calibration Scheme}
\label{subsec:res_selection}

Table~\ref{tab:selection} compares what each scheme shows on the calibration
part, on held-out folds and on the test part. The folds are five forward blocks
on Milan and entity-grouped partitions into three and two folds on the Testbed
and OpenIreland. On the calibration part the mean miss rate of every scheme
lies between $0.017$ and $0.050$ on every corpus. An in-sample compliance
criterion therefore admits all five schemes, and selection on it reduces to
minimum calibration duty. Scored on the same test data, each of the two
in-sample procedures produces a failure on one corpus. Single-seed selection
adopts the entity-standardised scheme on OpenIreland, which fails structurally
at test, whereas multi-seed selection adopts the pooled scheme on Milan, which
exceeds the tolerance systematically. Cross-validated selection avoids both
failures.

\begin{table}[!t]
\renewcommand{\arraystretch}{1.1}
\caption{Calibration Schemes. Mean Miss Rate ($R$) and Duty ($D$) Over Five
Seeds. Brackets Count the Seeds Whose Held-Out Miss Rate Lies Inside the
Tolerance, a Dash Marks Folds Not Defined on Every Seed, and the Adopted Scheme
Is Set in Bold}
\label{tab:selection}
\centering
\footnotesize
\setlength{\tabcolsep}{3.4pt}
\begin{tabular}{@{}llccccc@{}}
\toprule
 & & Cal. & \multicolumn{2}{c}{Held-out} & \multicolumn{2}{c}{Test} \\
\cmidrule(lr){3-3}\cmidrule(lr){4-5}\cmidrule(lr){6-7}
 & & $R$ & $R$ & $D$ & $R$ & $D$ \\
\midrule
\multirow{5}{*}{\rotatebox{90}{Milan}}
 & M0 & 0.050 & 0.032 [5] & 0.262 & 0.087 & 0.064 \\
 & \textbf{M1} & \textbf{0.049} & \textbf{0.056 [5]} & \textbf{0.234} & \textbf{0.063} & \textbf{0.094} \\
 & M2 & 0.046 & 0.028 [5] & 0.284 & 0.072 & 0.075 \\
 & M3 & 0.045 & 0.044 [5] & 0.264 & 0.069 & 0.087 \\
 & M4 & 0.050 & 0.032 [5] & 0.268 & 0.094 & 0.060 \\
\midrule
\multirow{3}{*}{\rotatebox{90}{Irish5G}}
 & \textbf{M0} & \textbf{0.038} & -- & -- & \textbf{0.068} & \textbf{0.135} \\
 & M1 & 0.050 & -- & -- & 0.067 & 0.134 \\
 & M4 & 0.038 & -- & -- & 0.045 & 0.524 \\
\midrule
\multirow{3}{*}{\rotatebox{90}{OpenIr.}}
 & M0 & 0.049 & 0.066 [3] & 0.119 & 0.033 & 0.122 \\
 & \textbf{M1} & \textbf{0.045} & \textbf{0.057 [4]} & \textbf{0.117} & \textbf{0.044} & \textbf{0.106} \\
 & M4 & 0.049 & 0.189 [3] & 0.111 & 0.209 & 0.114 \\
\midrule
\multirow{5}{*}{\rotatebox{90}{Testbed}}
 & \textbf{M0} & \textbf{0.041} & \textbf{0.043 [5]} & \textbf{0.625} & \textbf{0.107} & \textbf{0.555} \\
 & M1 & 0.050 & 0.048 [5] & 0.649 & 0.047 & 0.731 \\
 & M2 & 0.017 & -- & -- & 0.066 & 0.651 \\
 & M3 & 0.039 & -- & -- & 0.044 & 0.672 \\
 & M4 & 0.041 & 0.042 [5] & 0.688 & 0.049 & 0.683 \\
\bottomrule
\end{tabular}
\end{table}

Cross-validation proves decisive on OpenIreland, where the entity-standardised
scheme is excluded by both the silent-entity and the compliance conditions. In
one seed a held-out fold is silent, and the held-out event FNR reaches $0.667$,
or $9.9$ times the tolerance. The compliance condition is met on three of five
seeds only. The pooled scheme also meets it on three, which leaves online
calibration as the single admissible scheme. At test the rejected scheme misses
$93.8\%$ of the events of one seed at a duty of $0.014$. On that seed, $777$ of
the $778$ misses fall in two runs whose standardised score never reaches the
threshold ($1.95$ and $1.09$ against $2.28$). The cause is that per-entity
standardisation removes the level differences between entities on which a
pooled threshold operates.

The adopted configuration is online calibration on Milan and OpenIreland and
pooled calibration on Irish5G and the Testbed. On Milan the selected online
scheme is the only one of the five whose test miss rate lies within the
tolerance. No entity-conditional scheme survives on any corpus, which is why
the framework treats the scheme as selected and not as assumed. On Irish5G one
seed holds $55$ calibration events, below the declared minimum of $30$ per
held-out fold at any fold count. The scheme is therefore selected with a
single-seed procedure and is reported accordingly. The choice is immaterial
there, since the pooled and the online scheme yield the same test operating
point.

\subsection{Joint Certification and Dependence Structure}
\label{subsec:res_cert}

\begin{figure}[!t]
\centering
\includegraphics[width=0.60\linewidth]{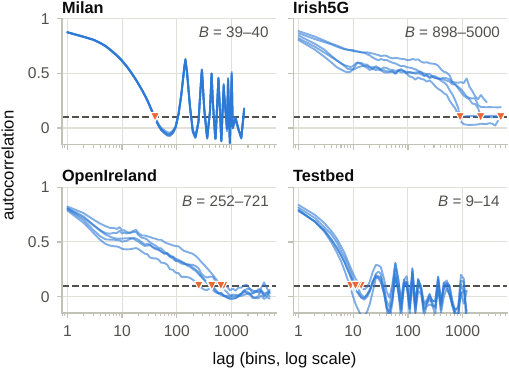}
\caption{Within-entity autocorrelation of the calibration alarm indicator under
the adopted scheme, with one curve per seed. Triangles mark the block length
$B$, the first lag below $0.1$ (dashed).}
\label{fig:dependence}
\end{figure}

Fig.~\ref{fig:dependence} shows why the unit of the duty test matters. The
alarm indicator is strongly autocorrelated on every corpus, with a lag-one
autocorrelation of $0.78$--$0.89$. The block length $B$ is $39$--$40$ bins on
Milan, $9$--$14$ on the Testbed, $252$--$721$ on OpenIreland and $898$ to the
search limit of $5000$ on Irish5G. Once the block length is taken from that
autocorrelation, the number of effective calibration units falls by one to
three orders of magnitude relative to the number of bins
(Table~\ref{tab:dependence}). Between $13$ and $9200$ blocks remain, against
$5.6\times10^{3}$ to $3.6\times10^{5}$ calibration bins.

Certifying the alarm budget alongside the miss rate succeeds in $12$ of the
$20$ corpus--seed combinations. On Milan the surviving set holds $489$--$525$
configurations on every seed at the nominal risk level. On the first seed the
set is unchanged under whole-cell units, so the daily periodicity visible in
Fig.~\ref{fig:dependence} does not alter that certificate. On OpenIreland four
seeds admit $105$--$393$ configurations at $\alpha_{\mathrm{J}}=0.10$ and a
budget of twice the calibration reference occupancy, and on the Testbed three
seeds admit $6$--$81$ configurations at $\alpha_{\mathrm{J}}=0.20$. At test,
every certified policy respects its duty budget. The realised miss rate lies
within the tolerance of its level in eleven of the twelve cases, in line with
the confidence level $1-\delta$. On the first seed the realised miss rates are
$0.0489$ on Milan and $0.0729$ on the Testbed, against levels of $0.05$ and
$0.20$. The levels differ across corpora because the search of
Section~\ref{subsec:joint} stops at the first pair that admits a survivor.

Where a certificate exists, its additional duty cost is modest. The certified
policies reduce the notification starts by $33$--$65\%$ at a duty increase of
at most $18\%$. On Milan the starts fall by $42\%$, from $0.063$ to $0.036$ per
hour, and on OpenIreland from $570$ to $325$ per hour. On Irish5G the block
length exceeds the median trace length on four of the five seeds, so a block is
a trace and the effective sample is reduced to $13$--$31$ units. No certificate
is issued there, because the configuration set is non-empty only under the
per-bin statistic. The limiting factor is therefore certificate availability
and not its duty cost.

\begin{table}[!t]
\renewcommand{\arraystretch}{1.12}
\caption{Dependence Structure and Joint Certification Over Five Seeds. Blocks
Is the Effective Number of Calibration Units, and Certified Counts the Seeds
With a Non-Empty Surviving Set}
\label{tab:dependence}
\centering
\footnotesize
\begin{tabular}{@{}lcccc@{}}
\toprule
Corpus & $B$ & Blocks & $(\alpha_{\mathrm{J}},\beta)$ & Certified \\
\midrule
Milan & 39--40 & 9000--9200 & (0.05, 0.93) & 5 of 5 \\
Irish5G & 898--5000 & 13--31 & (0.10, 0.20) & 0 of 5 \\
OpenIreland & 252--721 & 165--446 & (0.10, 0.24) & 4 of 5 \\
Testbed & 9--14 & 402--838 & (0.20, 0.76) & 3 of 5 \\
\bottomrule
\end{tabular}
\end{table}

\subsection{Layer Observability, Actionability and Footprint}
\label{subsec:res_testbed}

The Testbed answers a question that the public corpora cannot, namely how much
predictive skill each observability layer carries. Calendar features are
removed and all feature sets are scored on identical bins. The pinball loss
with application-layer features is lower than with RAN features by $0.389$,
$0.349$ and $0.485$\,ms at $h=5$, $10$ and $20$, with intervals excluding zero.
Their union is indistinguishable from the application layer alone. This is not
a sampling-rate artefact. With application features sampled at the RAN instants
and forward-filled identically, the RAN set remains worse by $0.444$, $0.458$
and $0.446$\,ms. On the decision-relevant metric, the event FNR difference
between the two layers is $-0.010$, $+0.003$ and $+0.008$, with intervals
including zero. Application-layer history predicts the delay value better but
carries no measurable advantage in anticipating the event, so an assurance
function without RAN counter access is not disabled.

The actionability analysis is an offline upper bound obtained without applying
a control action. In a single-seed replay with the regression score and online
calibration at $h=5$, $80$ of $96$ events are preceded by an alarm with a
median lead of $5$\,s, the length of the window. With the actuator reach held
at $20$\,s, the fraction of violating bins that an actuation window could cover
is $0.962$, $0.895$ and $0.789$ at reaction times of $1$, $5$ and $10$\,s. This
bound is read alongside its alarm cost of $24.7$ alarm bins per minute, of
which $85.2\%$ are false.

Compute latency, measured on one CPU core, is negligible relative to the bin
periods. With the onset classifier in the inference path, per-bin prediction
costs $3.9$--$10.4\,\mu$s when amortised over batches of $20\,000$ rows. The
online conformal update costs \mbox{$0.2$--$9.7\,\mu$s}. Against bin periods of
$0.1$\,s to $10$\,min, this leaves a margin of four to nine orders of
magnitude. Peak memory of $0.5$--$5.4$\,GB is incurred only in offline fitting.

\subsection{Discussion and Limitations}
\label{subsec:res_limits}

Three practical implications follow. First, an operating point is characterised
by three numbers, namely the miss rate, the duty relative to $D^{\star}$ and
the chance level. A miss rate close to chance indicates that the nominal level
was met through occupancy. Second, online calibration is the safer default
wherever violations are measured after the fact, which holds for every SLO used
here. Its long-run miss rate does not rest on exchangeability, and
\eqref{eq:identity} makes its finite-window excess auditable from the threshold
log. Third, the number of blocks should be computed before a joint certificate
is attempted. With tens of blocks no tight budget can be tested. Once a block
spans an entity, more entities are required, not more bins.

Four conditions bound the scope of these results. Milan is load derived from
call detail records of a production network and not 5G radio data. The Testbed
comprises one link, one UE and one cell over six hours, the scale at which its
alarm-layer result applies. The risk bound holds in expectation over the
calibration draw and not per test split. The duty certificates hold under the
block dependence model, and on the Testbed they are issued at a risk level four
times the nominal one.

\section{Conclusion}
\label{sec:conclusion}

This paper treated the conversion of a KPI forecast into an operational alarm
as the object to be calibrated. CALM calibrates on service-level violation
events, scores their onset directly, selects its conformal calibration scheme
by held-out scoring inside the calibration split, and reports the alarm duty
cycle against the occupancy of a full-knowledge oracle. On three of four
heterogeneous corpora the resulting layer stays within its miss-rate tolerance
while reducing time-in-alarm by $56.3\%$ to $89.7\%$. On the fourth the gain is
not significant.

Two findings generalise beyond this framework. First, pooled batch calibration
exceeds its miss-rate tolerance structurally under label shift, by $10.9$
standard errors on the production corpus, and online calibration restores
compliance. The choice between the two is therefore a validity question on
operational telemetry and not an efficiency one. Second, conformal calibration
schemes cannot be ranked on the split used to fit their thresholds. Both
in-sample procedures selected a scheme that failed out of sample, and held-out
scoring with deployment-matched folds resolved the failure.

One mechanism underlies both findings. Entity heterogeneity and within-entity
correlation reduce the number of independent calibration units by one to three
orders of magnitude relative to the number of bins. This reduction also
determines where the alarm budget can be certified jointly with the miss rate.
The layer observes and advises, and the actuation policy that consumes its
alarms remains a deployment choice.

\bibliographystyle{IEEEtran}
\bibliography{references}

\end{document}